\documentclass[aps,twocolumn,preprintnumbers,amsmath,amssymb,superscriptaddress,floatfix,nofootinbib]{revtex4}

\usepackage{lipsum}
\usepackage{graphicx}
\usepackage{epsfig}
\usepackage{epstopdf}
\usepackage{hyperref}
\usepackage{amsmath}
\usepackage{amsfonts}
\usepackage{amssymb}

\begin{document}

\title{Predictions for $\eta_c \to \eta \pi^+ \pi^-$ producing $f_0(500)$, $f_0(980)$ and $a_0(980)$}

\author{V.~R.~Debastiani}
\email{vinicius.rodrigues@ific.uv.es}
\affiliation{Institute of Modern Physics, Chinese Academy of
Sciences, Lanzhou 730000, China}
\affiliation{Departamento de
F\'{\i}sica Te\'orica and IFIC, Centro Mixto Universidad de
Valencia-CSIC Institutos de Investigaci\'on de Paterna, Aptdo.
22085, 46071 Valencia, Spain}

\author{Wei-Hong~Liang}
\email{liangwh@gxnu.edu.cn}
\affiliation{Department of Physics, Guangxi Normal University,
Guilin 541004, China}

\author{Ju-Jun~Xie}
\email{xiejujun@impcas.ac.cn}
\affiliation{Institute of Modern Physics, Chinese Academy of
Sciences, Lanzhou 730000, China}

\author{E.~Oset}
\email{oset@ific.uv.es}
\affiliation{Institute of Modern Physics, Chinese Academy of
Sciences, Lanzhou 730000, China}
\affiliation{Departamento de
F\'{\i}sica Te\'orica and IFIC, Centro Mixto Universidad de
Valencia-CSIC Institutos de Investigaci\'on de Paterna, Aptdo.
22085, 46071 Valencia, Spain}

\date{\today}

\begin{abstract}
We perform calculations for the $\eta_c \to \eta \pi^+ \pi^-$ decay using elements of SU(3) symmetry to see the weight of different trios of pseudoscalars produced in this decay, prior to the final state interaction of the mesons. After that, the interaction of pairs of mesons, leading finally to $\eta \pi^+ \pi^-$, is done using the chiral unitary approach. We evaluate the $\pi^+ \pi^-$ and $\pi \eta$ mass distributions and find large and clear signals for $f_0(500)$, $f_0(980)$ and $a_0(980)$ excitation. The reaction is similar to the $\chi_{c1} \to \eta \pi^+ \pi^-$, which has been recently measured at BESIII and its implementation and comparison with these predictions will be very valuable to shed light on the nature of the low mass scalar mesons.\\

\end{abstract}

\maketitle

\section{Introduction}

The sector of light scalar mesons has been a topic of intense discussions for years \cite{Klempt:2007cp,vanBeveren:1986ea,Tornqvist:1995ay,Fariborz:2009cq,Fariborz:2009wf}. Early discussions on their nature as $q \bar q$ or more complex objects have converged to accept that these states cannot be $q \bar q$ objects. An extensive updated discussion on the issue can be seen in the report \cite{Pelaez:2015qba}. The discussions in Ref. \cite{Pelaez:2015qba} reveal the large amount of empirical information favoring a dynamical picture in which the interaction of pseudoscalar mesons in coupled channels and constraints of unitarity generate scalar mesons, which would qualify as multichannel pseudoscalar-pseudoscalar molecular states. The successful picture incorporating the constraints of unitarity in coupled channels and the dynamics of the chiral Lagrangians \cite{Weinberg:1978kz,Gasser:1983yg,Bernard:1995dp} is known as the chiral unitary approach, and either using the inverse amplitude method \cite{Dobado:1996ps,Pelaez:2006nj,Oller:1998hw} or the coupled channels Bethe-Salpeter equations \cite{Oller:1997ti,Kaiser:1998fi,Locher:1997gr,Nieves:1999bx}, the success in providing an accurate description in the different reactions where these resonances are produced is remarkable. Detailed reviews of such reactions can be seen in Ref. \cite{Oller:2000ma} and more recently in Ref. \cite{Oset:2016lyh}, in relation to $B$, $D$, $\Lambda_b$ and $\Lambda_c$ decays involving these resonances as dynamically generated. We note in passing that the chiral unitary approach does not implement crossing symmetry, which means that it cannot be used to obtain $\pi K \to \pi K$ from the $\pi \pi \to K \bar K$ amplitudes. In practice what one does is to unitarize $\pi \pi \to K \bar K$ and $\pi K \to \pi K$ in the physical region as independent reactions. This procedure leads to amplitudes in remarkable agreement with semiempirical studies using the Roy equation, where crossing is also implemented \cite{Ananthanarayan:2000ht,GarciaMartin:2011cn}.

Tetraquark pictures have also been invoked \cite{Jaffe,Fariborz:2009cq}, but the standard configurations chosen to account for the masses run into one or another problem in different reactions. A detailed discussion on this issue can be seen in section IV of Ref. \cite{Dias:2016gou}.

What ultimately sets the balance in favor of one or another theoretical picture is the power to provide an accurate explanation of multiple reactions, and in this sense there is nothing more convincing than making predictions for reactions not yet measured and having the predictions realized a posteriori by experiment. This is the aim of the present work where we make predictions for the decay $\eta_c \to \eta \pi^+ \pi^-$ looking into the invariant mass distributions of $\pi\pi$ and $\pi \eta$. In the distributions we find a very clear and strong signal for the $a_0(980)$, and also clearly seen but weaker signals for $f_0(500)$ and $f_0(980)$ excitation. We are confident on the results up to invariant masses of about 1200 MeV and propose the measurement of this reaction that can easily be implemented in BESIII.

There is a precedent for the $\eta_c \to \eta \pi^+ \pi^-$ reaction in the $\chi_{c1} \to \eta \pi^+ \pi^-$ decay, which has been measured at BESIII \cite{Kornicer:2016axs}. In this reaction one can see a neat $a_0(980)$ signal in the $\pi \eta$ mass distribution with its typical cusp shape and with very little background. On the other hand, in the $\pi^+ \pi^-$ mass spectrum one sees a very clear peak for the $f_0(500)$ and a smaller, but visible peak for the $f_0(980)$. The $\pi^+ \pi^-$ spectrum also shows a pronounced signal for the $f_2(1270)$ excitation. A theoretical study for this reaction using the chiral unitary approach was done in Ref. \cite{Liang:2016hmr} and a good reproduction of the shapes and relative strengths of the invariant mass distributions was obtained up to about 1200 MeV, the present limit of applicability of the chiral unitary approach in the interaction of pseudoscalar mesons.

The $\eta_c \to \eta \pi^+ \pi^-$ has many things in common to the $\chi_{c1} \to \eta \pi^+ \pi^-$, but also differences. The $\chi_{c1}$ has quantum numbers $I^G(J^{PC})=0^+(1^{++})$, the $\eta_c$ has $0^+(0^{-+})$. In the $\chi_{c1} \to \eta \pi^+ \pi^-$, if the $\pi^+ \pi^-$ is in $S$-wave to create the $f_0(500)$ and $f_0(980)$, the $\eta$ must be in $P$-wave to conserve angular momentum and parity. In the $\eta_c$ decay the process can proceed in $S$-wave. Concerning the $f_2(1270)$ excitation, in the $\chi_{c1} \to \eta \pi^+ \pi^-$ reaction, the same process with $\eta$ in $P$-wave, and the $\pi^+ \pi^-$ in $D$-wave, can produce the resonance. In the $\eta_c \to \eta \pi^+ \pi^-$ we will need a $D$-wave for $\eta$ in the production vertex, in addition to the internal $D$-wave of $\pi^+ \pi^-$. This mechanism should be suppressed versus the one of $f_0(500)$ or $f_0(980)$ production and then the signals for the scalar mesons would be cleaner than those in the $\chi_{c1} \to \eta \pi^+ \pi^-$ reaction. With this perspective we perform the calculations and make predictions for the reaction. In the absence of the $f_2(1270)$ excitation we also make predictions for the background. Our limitations to the range below 1200 MeV for the energies of the interacting meson pairs, induce uncertainties on the background, but we can show that these uncertainties are small in the region of $\pi^+ \pi^-$ or $\pi \eta$ invariant masses below 1200 MeV, thus making our predictions really solid. With these results and clear predictions, we can only encourage the performance of the experiment which is easily implementable at BESIII.

\section{Formalism}

We follow closely the work of Ref. \cite{Liang:2016hmr} with the particular differences of this case. As commented before, the process proceeds in $S$-wave. We consider that $\eta_c$ behaves as a singlet of SU(3), since it does not have $u$, $d$, $s$ quarks and, hence, we must construct an SU(3) singlet with the product of three pseudoscalars. For this purpose we write the $\phi$ SU(3) matrix corresponding to $q \bar q$, including the mixing of $\eta$, $\eta'$ with
\begin{align}
\begin{aligned}
\eta &= \cos\theta_P \, \eta_8 - \sin\theta_P \, \eta_1 \, , \\
\eta' &= \sin\theta_P \, \eta_8 + \cos\theta_P \, \eta_1 \, ,
\end{aligned}
\end{align}
with $\sin\theta_P = -1/3$, which is a standard choice \cite{Bramon:1992kr}. A more recent determination of this angle from fits to world data is done in Ref. \cite{Ambrosino:2009sc} with $\theta_P = -14.34 \, ^\circ$. The dominant $\eta$ component going with $\cos\theta_P$ only changes by 3\% by taking $\theta_P = -14.34 \, ^\circ$ or $\sin\theta_P = -1/3$, and then we choose $\sin\theta_P = -1/3$ which leads to a convenient form of the $\phi$ matrix, given by

\begin{widetext}
\begin{equation}\label{eq:phi}
\phi \equiv \left(
           \begin{array}{ccc}
             \frac{1}{\sqrt{2}}\pi^0 + \frac{1}{\sqrt{3}}\eta + \frac{1}{\sqrt{6}}\eta' & \pi^+ & K^+ \\
             \pi^- & -\frac{1}{\sqrt{2}}\pi^0 + \frac{1}{\sqrt{3}}\eta + \frac{1}{\sqrt{6}}\eta' & K^0 \\
            K^- & \bar{K}^0 & -\frac{1}{\sqrt{3}}\eta + \sqrt{\frac{2}{3}}\eta' \\
           \end{array}
         \right).
\end{equation}
\end{widetext}

We can write three independent SU(3) invariants with three $\phi$ matrices: ${\rm Trace}(\phi\phi\phi)$, ${\rm Trace}(\phi){\rm Trace}(\phi\phi)$, $[{\rm Trace}(\phi)]^3$. They are written in terms of the mesons as
\begin{align}\label{eq:phiphiphi}
\nonumber {\rm Trace} (\phi \phi \phi) &= 2\sqrt{3} \eta \pi^+ \pi^-
   + \sqrt{3} \eta \pi^0 \pi^0
   + \frac{\sqrt{3}}{9} \eta \eta \eta \\
&   + 3 \pi^+ K^0 K^- + 3 \pi^- K^+ \bar K^{0},
\end{align}
\begin{align}\label{eq:TphiTphiphi}
\nonumber {\rm Trace}(\phi) {\rm Trace}(\phi \phi) &=
\frac{\sqrt{3}}{3} \,\eta \,
( 2\pi^+ \pi^-
+ \pi^0 \pi^0
+ 2K^+ K^-\\
&+ 2K^0 \bar K^0
+ \eta \eta),
\end{align}
\begin{equation}\label{eq:Tphi3}
[{\rm Trace}(\phi)]^3 = \frac{\sqrt{3}}{9}\, \eta \eta \eta.
\end{equation}

In Eqs. (\ref{eq:phiphiphi}), (\ref{eq:TphiTphiphi}), (\ref{eq:Tphi3}) we have removed the $\eta'$ components, which play only a marginal role in the building of the $f_0(500)$, $f_0(980)$, $a_0(980)$ resonances, because of its large mass and small couplings. We have also eliminated other terms like $\pi^0 K^+ K^-$ because upon final state interaction of any pair, as we do here, we never have the $\eta \pi^+ \pi^-$ combination which is measured experimentally.

These expressions give us the relative weights in which trios of pseudoscalars are produced from $\eta_c$ decay in a first step, prior to the final state interaction of these mesons.

\begin{figure}[tb]\centering
\includegraphics[width=0.5\textwidth]{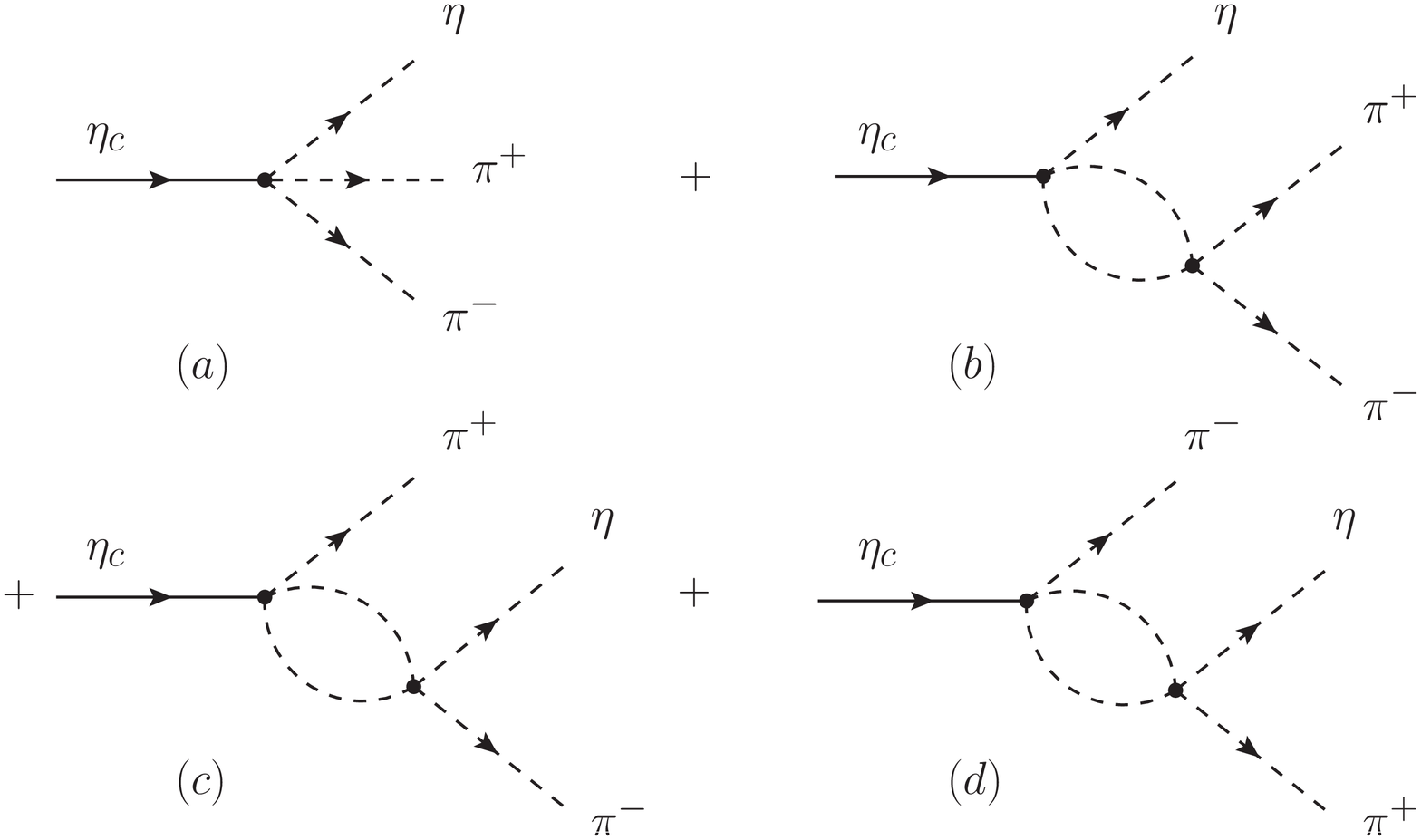}
\caption{Diagrams involved in the $\eta_c \to \eta \pi^+ \pi^-$ reaction including final state interaction of pairs of mesons.}
\label{fig:FeynmanDiag}
\end{figure}

The next step is to allow them to interact. By letting all possible pairs to interact and make transitions, and isolating the final $\eta \pi^+ \pi^-$ channel, the diagrams to be considered are given in Fig. \ref{fig:FeynmanDiag}.

In the loops of Fig. \ref{fig:FeynmanDiag} we show all pairs allowed by Eqs. (\ref{eq:phiphiphi}), (\ref{eq:TphiTphiphi}), (\ref{eq:Tphi3}) that can give rise to the considered final state. Then the amplitude that sums all terms is given by
\begin{equation}\label{eq:t}
t = t_{tree} + t_{\eta} + t_{\pi^+} + t_{\pi^-},
\end{equation}
where the tree-level amplitude is
\begin{equation}\label{eq:t_tree}
t_{tree} = V_p h_{\eta \pi^+ \pi^-},\\
\end{equation}
and the first transition amplitude is
\begin{equation}\label{eq:t_eta}
t_{\eta} = V_p \sum_i h_i S_i G_i(M_{\rm inv}(\pi^+ \pi^-)) t_{i,\pi^+ \pi^-}(M_{\rm inv}(\pi^+ \pi^-)),
\end{equation}
where $V_p$ is a constant coefficient, common to all four terms, that accounts for the matrix element of the tree-level $\eta_c \to 3$ mesons transition, up to $h_i$ coefficients, which are the factors multiplying each combination of three mesons in Eqs. (\ref{eq:phiphiphi}), (\ref{eq:TphiTphiphi}), (\ref{eq:Tphi3}). The values of $h_i$ for the term $ {\rm Trace}(\phi\phi\phi)$ are
\begin{align}
\begin{aligned}
h_{\eta \pi^+ \pi^-} = 2\sqrt{3}, ~~~~
h_{\eta \pi^0 \pi^0} = \sqrt{3},\\
h_{\eta \eta \eta} = \frac{\sqrt{3}}{9}, ~~~~
h_{\pi^+ K^0 K^-} = h_{\pi^- K^+ \bar K^0} = 3.
\end{aligned}
\end{align}

In this case, in Eq. (\ref{eq:t_eta}) only $h_{\eta \pi^+ \pi^-}$, $h_{\eta \pi^0 \pi^0}$ and $h_{\eta \eta \eta}$ contribute. The function $G_i$ is the loop function of the two intermediate mesons and $t_{i,\pi^+ \pi^-}$ is the transition matrix element from the state $i$ to $\pi^+ \pi^-$. The $G_i$ and $t_{i,\pi^+ \pi^-}$ functions, depending on the invariant masses of $\pi^+ \pi^-$, $M_{\rm inv}(\pi^+ \pi^-)$, are taken from the chiral unitary approach, and we follow Refs. \cite{Oller:1997ti,Liang:2014tia,Xie:2014tma,Liang:2014ama}. The factor $S_i$ is a symmetry factor to account for identical particles,
\begin{equation}\label{eq:S_factor}
  S_{\pi^0 \pi^0}=2! \frac{1}{2} ~~({\rm for}~~ \pi^0 \pi^0),~~ S_{\eta \eta}=3! \frac{1}{2}~~ ({\rm for}~~ \eta \eta).
\end{equation}

Similarly, we have
\begin{equation}
\label{eq:t_Pi+} t_{\pi^+} = V_p \sum_i h_i S_i G_i(M_{\rm inv}(\pi^- \eta)) t_{i,\pi^- \eta}(M_{\rm inv}(\pi^- \eta)),
\end{equation}
where in the sum over $i$ we have the states $\pi^- \eta$ and $K^0 K^-$, and
\begin{equation}
\label{eq:t_Pi-} t_{\pi^-} = V_p \sum_i h_i S_i G_i(M_{\rm inv}(\pi^+ \eta)) t_{i,\pi^+ \eta}(M_{\rm inv}(\pi^+ \eta)),
\end{equation}
where now in $i$ we have $\pi^+ \eta$ and $K^+ \bar K^0$.

We can do the same for the ${\rm Trace}(\phi) {\rm Trace}(\phi \phi)$ and $[{\rm Trace}(\phi)]^3$, and we would have an indetermination in the relative weight of the three different terms. At this point we take advantage to complement the information given in Ref. \cite{Liang:2016hmr} for the $\chi_{c1} \to \eta \pi^+ \pi^-$ reaction. The flavor content in the present reaction and in the $\chi_{c1} \to \eta \pi^+ \pi^-$ is identical, only a $P$-wave vertex appears in this latter reaction, while here it proceeds via $S$-wave.
\begin{figure}[tb]\centering
\includegraphics[width=0.55\textwidth]{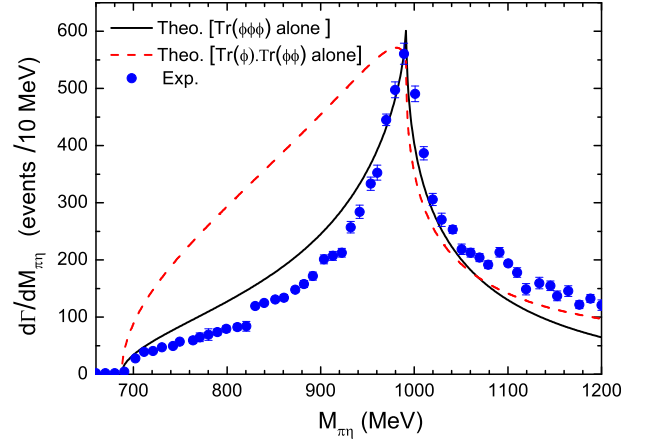}
\caption{Results for the $\pi\eta$ mass distribution in the $\chi_{c1} \to \eta \pi^+ \pi^-$ reaction. Data from Ref. \cite{Kornicer:2016axs}. Solid curve: results from Ref. \cite{Liang:2016hmr} using ${\rm Trace}(\phi\phi\phi)$. Dashed line: results using ${\rm Trace}(\phi){\rm Trace}(\phi\phi)$ normalized to the peak of the distribution.}
\label{fig:chic1a0}
\end{figure}
\begin{figure}[tb]\centering
\includegraphics[width=0.55\textwidth]{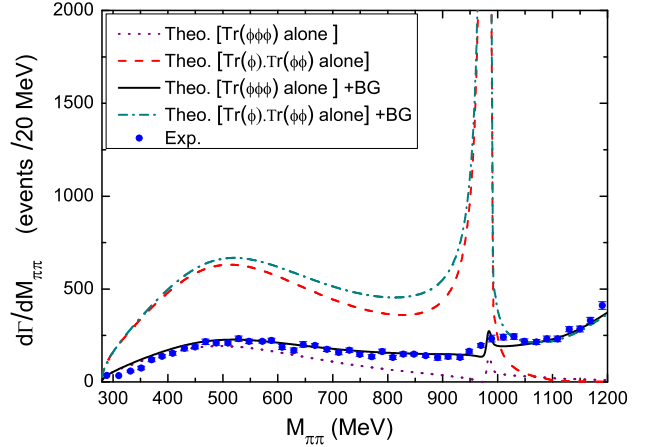}
\caption{Results for the $\pi^+ \pi^-$ distribution in the $\chi_{c1} \to \eta \pi^+ \pi^-$ reaction. Data from Ref. \cite{Kornicer:2016axs}. Dotted and solid lines: results from Ref. \cite{Liang:2016hmr} using ${\rm Trace}(\phi\phi\phi)$, with and without background contribution. Dash-dotted and dashed lines: results using ${\rm Trace}(\phi){\rm Trace}(\phi\phi)$, with and without background.}
\label{fig:chic1f0}
\end{figure}

We show in Figs. \ref{fig:chic1a0} and \ref{fig:chic1f0}, the results of Ref. \cite{Liang:2016hmr} using only ${\rm Trace} (\phi\phi\phi)$ or ${\rm Trace}(\phi){\rm Trace}(\phi\phi)$. The results have been normalized in both cases to the peak of the $\pi\eta$ invariant mass distribution in Fig. \ref{fig:chic1a0}.

We observe that the shape for the case of ${\rm Trace}(\phi){\rm Trace}(\phi\phi)$ is completely off from experiment \cite{Kornicer:2016axs}. Similarly, the strength of the $\pi^+ \pi^-$ distribution is also much bigger than experiment and it produces a huge $f_0(980)$ peak, in total disagreement with experiment. We have also tried different linear combinations of ${\rm Trace}(\phi\phi\phi)$, ${\rm Trace}(\phi){\rm Trace}(\phi\phi)$ and $[{\rm Trace}(\phi)]^3$, concluding that the best reproduction of the data is obtained with the term ${\rm Trace}(\phi\phi\phi)$ alone, which is also more symmetrical in the three mesons. The role of the term $[{\rm Trace}(\phi)]^3$, involving only $\eta$ mesons, is negligible for values of its strength of the order of magnitude of the one of ${\rm Trace}(\phi\phi\phi)$. In view of these results, we make predictions for the $\eta_c \to \eta \pi^+ \pi^-$ with only the term ${\rm Trace}(\phi\phi\phi)$.

We take as reference the $\pi^+ \pi^-$ and $\pi^+ \eta$ invariant masses and write the double differential mass distribution for three-body decays \cite{PDG}
\begin{align}\label{eq:d2Gamma}
\begin{aligned}
&\frac{\rm d^2 \Gamma}{\rm d M_{\rm inv}(\pi^+ \pi^-) d M_{\rm inv}(\pi^+ \eta)}\\
&= \frac{1}{(2\pi)^3} \frac{1}{8 M_{\eta_c}^3} M_{\rm inv}(\pi^+ \pi^-) M_{\rm inv}(\pi^+ \eta) |t|^2.
\end{aligned}
\end{align}

From this formula we obtain $\frac{\rm d\Gamma}{\rm d M_{\rm inv}(\pi^+ \pi^-)}$ and $\frac{\rm d\Gamma}{\rm d M_{\rm inv}(\pi^+ \eta)}$ by integrating over the other invariant mass. By labeling 1, 2, 3 to the $\eta$, $\pi^+$, $\pi^-$ particles, respectively, if we integrate over $M_{23}$, the limits of integration are given in Ref. \cite{PDG} (alternative, equivalent, expressions can be obtained from Ref. \cite{Byckling:1973}). These limits are
\begin{align}\label{eq:limM23}
\begin{aligned}
(M_{23}^2)_{max} &= (E_2^* + E_3^*)^2 \\
&- (\sqrt{{E_2^*}^2-m_2^2}-\sqrt{{E_3^*}^2-m_3^2}~)^2,\\
(M_{23}^2)_{min} &= (E_2^* + E_3^*)^2 \\
&- (\sqrt{{E_2^*}^2-m_2^2}+\sqrt{{E_3^*}^2-m_3^2}~)^2,
\end{aligned}
\end{align}
where
\begin{align}\label{eq:ElimM23}
\begin{aligned}
E_2^* &= (M_{12}^2 - m_1^2 + m_2^2)/2M_{12},\\
E_3^* &= (M_{\eta_c}^2 -M_{12}^2 -m_3^2)/2M_{12}.
\end{aligned}
\end{align}

If we integrate over $M_{12}$, the limits of integration are
\begin{align}\label{eq:limM12}
\begin{aligned}
(M_{12}^2)_{max} &= (E_2^{*'} + E_1^{*'})^2 \\
&- (\sqrt{{E_2^{*'}}^2-m_2^2}-\sqrt{{E_1^{*'}}^2-m_1^2}~)^2,\\
(M_{12}^2)_{min} &= (E_2^{*'} + E_1^{*'})^2 \\
&- (\sqrt{{E_2^{*'}}^2-m_2^2}+\sqrt{{E_1^{*'}}^2-m_1^2}~)^2,\\
\end{aligned}
\end{align}
where
\begin{align}\label{eq:ElimM12}
\begin{aligned}
E_2^{*'} &= (M_{23}^2 - m_3^2 + m_2^2)/2M_{23},\\
E_1^{*'} &= (M_{\eta_c}^2 -M_{23}^2 -m_1^2)/2M_{23}.
\end{aligned}
\end{align}

Since we take the $\pi^+ \pi^-$ and $\pi^+ \eta$ invariant masses as variables, we must note that $t_{\pi^+}$ depends on the $\pi^- \eta$ invariant mass, $M_{13}$. However, this mass is given in terms of the other two variables since one has \cite{PDG}
\begin{equation}
M^2_{13} = M^2_{\eta_c} + 2 m^2_{\pi} + m^2_{\eta} - M^2_{12} - M^2_{23}.
\end{equation}

\begin{figure}[tb]\centering
\includegraphics[width=0.5\textwidth]{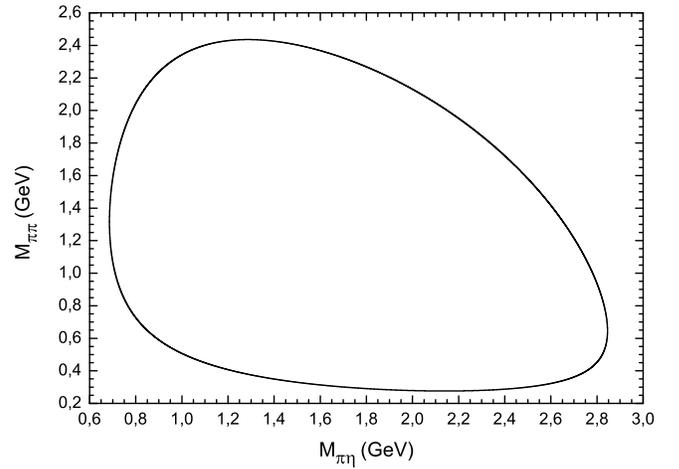}
\caption{Dalitz Plot for $\eta_c \to \eta \pi^+ \pi^-$, in the $\pi\eta$ and $\pi\pi$ masses.}
\label{fig:Dalitz}
\end{figure}

\section{Results}

For simplicity, we will refer to $M_{\rm inv}(\pi^+ \pi^-)$ and $M_{\rm inv}(\pi^+ \eta)$ as $M_{\pi\pi}$ and $M_{\pi\eta}$ respectively.
In Fig. \ref{fig:Dalitz} we show the Dalitz plot for $M_{\pi\pi}$ and $M_{\pi\eta}$ in the $\eta_c \to \eta \pi^+ \pi^-$ decay.
We are interested in $\frac{\rm d\Gamma}{\rm d M_{\rm inv}(\pi^+ \pi^-)}$ and $\frac{\rm d\Gamma}{\rm d M_{\rm inv}(\pi^+ \eta)}$ in the region of $f_0(500)$, $f_0(980)$ and $a_0(980)$. If we take $M_{\pi\eta}\sim~1000$ MeV we see that $M_{\pi\pi}$ goes from 500-2300 MeV, but the range is similar for values of $M_{\pi\eta}$ up to 2200 MeV.
This means that the strength of the $\pi\pi$ distribution will be spread along a wide range of $M_{\pi\eta}$ and we expect roughly a background following phase space.
At $M_{\pi\eta}\sim~750$ MeV the range of $M_{\pi\pi}$ is reduced to 800-1700 MeV and we can expect to obtain contribution from the $M_{\pi\pi}\sim980$ MeV region, which we have under control. Altogether we might anticipate that the background below the $a_0(980)$ peak will be moderate and controllable.

If we now fix $M_{\pi\pi}$ in 500-1000 MeV, the range of $M_{\pi\eta}$ is large and we should expect a background evenly distributed according to phase space. However, for $M_{\pi\pi}\sim400$ MeV the range of $M_{\pi\eta}$ begins at 1200 MeV, thus for these energies we will not have contribution from the large peak of the $a_0(980)$ and the background will be small.

In order to evaluate the differential mass distributions we must bear in mind that the chiral unitary approach that we use only makes reliable predictions up to 1100-1200 MeV. One should not use the model for higher invariant masses. With this perspective we will have to admit uncertainties in the mass distributions, particularly at invariant masses higher than 1200 MeV which are a large part of the Dalitz plot. Yet, we are only interested in the region of invariant masses below 1200 MeV both in $M_{\pi\pi}$ and $M_{\pi\eta}$ and it is just there where we would like to know uncertainties of our model. For that purpose we take the following prescription: we evaluate $Gt(M_{\rm inv})$ combinations up to $M_{\rm inv}=M_{\rm cut}$. From there on, we multiply $Gt$ by a smooth factor to make it gradually decrease at large $M_{\rm inv}$. Thus we take
\begin{equation}\label{Gt}
  Gt(M_{\rm inv}) = Gt(M_{\rm cut}){\rm e}^{-\alpha(M_{\rm inv}-M_{\rm cut})},
  ~~ {\rm for} ~~M_{\rm inv} > M_{\rm cut}.
\end{equation}

\begin{figure}[tb]\centering
\includegraphics[width=0.5\textwidth]{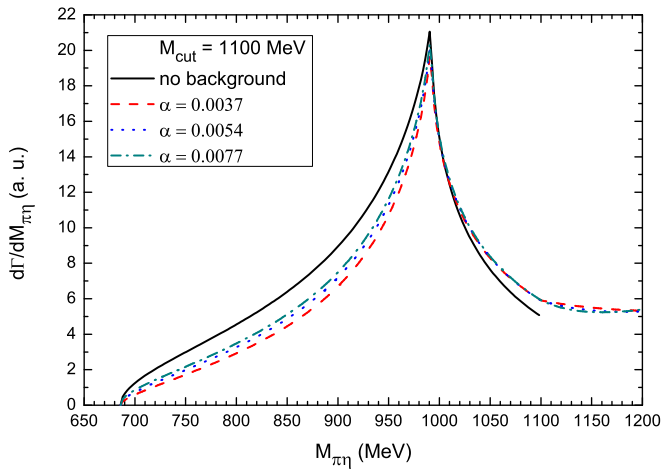}
\caption{(Color online) $\frac{\rm d\Gamma}{\rm d M_{\pi\eta}}$ as a function of $M_{\pi\eta}$ for $M_{\rm cut}=1100$ MeV and three different values of $\alpha$. See text for explanations.}
\label{1100a0}
\end{figure}

\begin{figure}[tb]\centering
\includegraphics[width=0.5\textwidth]{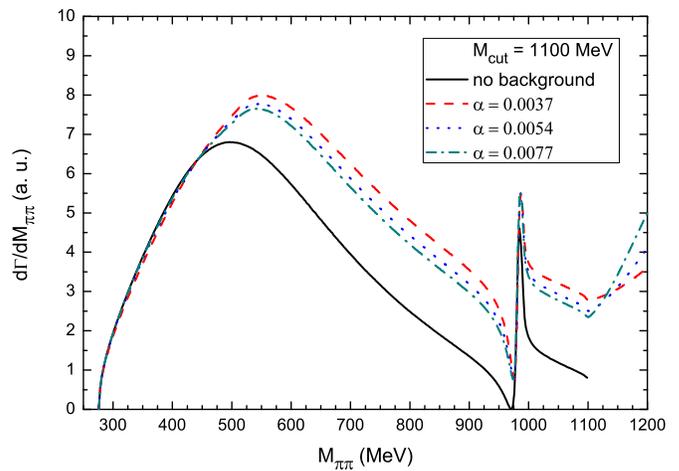}
\caption{(Color online) $\frac{\rm d\Gamma}{\rm d M_{\pi\pi}}$ as a function of $M_{\pi\pi}$ for $M_{\rm cut}=1100$ MeV and three different values of $\alpha$. See text for explanations.}
\label{1100f0}
\end{figure}

We take the value $M_{\rm cut}=$ 1100 MeV, with $\alpha=$ 0.0037 MeV$^{-1}$, 0.0054 MeV$^{-1}$ and 0.0077 MeV$^{-1}$, which reduce $Gt$ by about a factor 3, 5 and 10, respectively, at $M_{\rm cut} + 300$ MeV. We show the results in Fig. \ref{1100a0} for $\frac{\rm d\Gamma}{\rm d M_{\rm inv}(\pi^+ \eta)}$ and Fig. \ref{1100f0} for $\frac{\rm d\Gamma}{\rm d M_{\rm inv}(\pi^+ \pi^-)}$. The results taking $M_{\rm cut}=$ 1150 MeV are practically identical below 1200 MeV.

In Fig. \ref{1100a0} we show our results for $\frac{\rm d\Gamma}{\rm d M_{\rm inv}(\pi^+ \eta)}$ (the $\frac{\rm d\Gamma}{\rm d M_{\rm inv}(\pi^- \eta)}$ is identical). We see that below 1200 MeV, in the region of the $a_0(980)$, the uncertainties are very small, what makes the predictions in that region rather reliable. Since the amplitude $t$ of Eq. (\ref{eq:t}) sums coherently all terms, it is interesting to see what is mostly responsible for the peak. For this we keep in $t$ only the tree-level amplitude $t_{tree}$ and $t_{\pi^-}$, since $t_{\pi^-}$ is the term that contains the direct $M_{\rm inv}(\pi^+ \eta)$ dependence in $t_{i,\pi^+ \eta}(M_{\rm inv}(\pi^+ \eta))$. The result obtained with these two terms are shown in Fig. \ref{1100a0} by the solid line. This is what we call in the figure, ``no background''. We can see that the ``background'' created in that region by the other two terms, $t_{\pi^+}$ and $t_{\eta}$ is rather small. Yet, in the region from $M_{\pi\eta}=$ 700 MeV to 990 MeV, this ``background'' reduces a bit the contribution obtained by $t_{tree} + t_{\pi^-}$ only.

It is interesting to note that the ``no background'' prescription was taken in Ref. \cite{Liang:2016hmr}, and a smooth background was added incoherently to the $\pi\pi$ mass distributions in the $\chi_{c1} \to \eta \pi^+ \pi^-$, but not to the $\pi\eta$ mass distribution. The $a_0(980)$ mass distribution was in quite good agreement with experiment \cite{Kornicer:2016axs}, but was a bit higher in the $M_{\pi\eta}= 700-990$ MeV region, by an amount similar to the difference seen in Fig. \ref{1100a0} between the solid and other curves. The results obtained here could be easily translated there, with the consequent improvement of the agreement with the data. Similarly, at energies above 1000 MeV the ``background'' increases the ``no background'' curve, and this could also help the results of Ref. \cite{Liang:2016hmr} to get in better agreement with the data of Ref. \cite{Kornicer:2016axs}.

The strong cusp shape of the $a_0(980)$ and the small background, qualify this reaction, together with the $\chi_{c1} \to \eta \pi^+ \pi^-$, as the reaction where $a_0(980)$ shows up more strongly and more neatly.

In Fig. \ref{1100f0} we show the analogous results of Fig. \ref{1100a0} but for the $\frac{\rm d\Gamma}{\rm d M_{\rm inv}(\pi^+\pi^-)}$ mass distribution. We see that taking $M_{\rm cut}=$ 1100 MeV, there are uncertainties in the region of $M_{\pi\pi}>1200$ MeV for the different values of $\alpha$ chosen, but the uncertainties are much smaller in the region below 1200 MeV, what makes the predictions more solid.  It is more interesting to see that we observe a neat signal for the $f_0(500)$ and a much smaller, but clearly visible, signal for the $f_0(980)$. We also show the results with ``no background'' obtained taking for $t$ the sum $t_{tree} + t_{\eta}$, since in $t_{\eta}$ we have the terms $t_{i,\pi^+ \pi^-}(M_{\rm inv}(\pi^+ \pi^-))$. We can see that the ``background'' does not affect the mass distribution below 450 MeV, but gives a sizeable contribution from 550 MeV to 1200 MeV. Once more, in the $\chi_{c1} \to \eta \pi^+ \pi^-$ reaction studied in Ref. \cite{Liang:2016hmr}, where only the ``no background'' terms were considered, it was found that an ``empirical'' background of this size was needed to reproduce the data of Ref. \cite{Kornicer:2016axs}. Again, all these facts reinforce the reliability of the predictions made here.

The results obtained are shown in arbitrary units (the calculations are done taking a value of $V_p=100$), however, the relative weights for the $M_{\pi\pi}$ and $M_{\pi\eta}$ mass distributions of the figures are also predictions that can be tested in actual experiment. As we can see, the strength of the peak of the $a_0(980)$ is about three times the strength of the $f_0(500)$ peak.

We should note that in the region of $M_{\pi\pi}$ or $M_{\pi\eta}$ above 1200 MeV one should expect contribution from other resonances, not accounted for here. However, the small uncertainties of the spectrum below 1200 MeV due to the uncertainties above 1200 MeV, indicate that the corrections below 1200 MeV due to the contribution of higher energy resonances would still be small.

We can be more quantitative about this by looking at the amplitude analysis done in Ref. \cite{Kornicer:2016axs}. In Fig. 6 of that work, one can see contributions of $a_0(980)\pi$, $a_2(1320)\pi$, $a_2(1700)\pi$, $S_{K \bar K \to \pi \pi} \eta$, $S_{\pi \pi \to \pi \pi} \eta$, $f_2(1270)\eta$, $f_4(2050)\eta$. What is seen there is that all these terms (except for the $a_0(980)\pi$ itself) give a negligible background in the $a_0(980)\pi$ peak below 1200 MeV. On the other hand, the $\pi\pi$ distribution is dominated by the $S_{\pi \pi \to \pi \pi} \eta$ term (leading to the $f_0(500)$ peak) and the $S_{K \bar K \to \pi \pi} \eta$ term (leading to the $f_0(980)$ peak). All the other terms, except for the replica of the $a_0(980)\pi$ peak, give also negligible contribution in the $\pi\pi$ mass distribution below 1200 MeV. Only the $f_2(1270)\eta$ gives some small contribution around 1200 MeV, but we argued that here it should be suppressed. The replica of the $a_0(980)\pi$ peak in the $\pi\pi$ mass distribution we have here, and it is basically responsible for the differences that we have in Fig. \ref{1100f0} between the ``no background'' and the total contributions, similarly as to what is found in Ref. \cite{Liang:2016hmr}.

\section{Conclusions}

We have done a theoretical study of the $\eta_c \to \eta \pi^+ \pi^-$ decay paying attention to the final state interaction of the pairs of mesons. We evaluate  $\frac{\rm d\Gamma}{\rm d M_{\rm inv}(\pi^+ \pi^-)}$ and $\frac{\rm d\Gamma}{\rm d M_{\rm inv}(\pi^+ \eta)}$ and make predictions that should be confronted by a future experiment. The first step is to see the weight of the possible trios of mesons coming from $\eta_c$ decay, prior to any final state interaction, which is done assuming that $\eta_c$ is an SU(3) singlet and then using SU(3) symmetry in the trios of pseudoscalar mesons. We relied upon the results of the $\chi_{c1} \to \eta \pi^+ \pi^-$ reaction to support the fact that the ${\rm Trace}(\phi\phi\phi)$, most symmetric in the three fields, is the appropriate invariant to be used in the present reaction. After that, the interaction of all possible pairs in the trios (not only $\eta \pi^+ \pi^-$) are allowed to interact, leading to the final $\eta \pi^+ \pi^-$. The calculations are done using the chiral unitary approach for the interaction of mesons, which has a limit of applicability up to $M_{\rm inv}=1200$ MeV. We observe a large and clean signal for the $a_0(980)$ in the $\pi \eta$ mass distribution, and a relatively large signal for $f_0(500)$  and a smaller one for $f_0(980)$ in the $\pi^+ \pi^-$ mass distribution.

Given our ignorance above 1200 MeV, we kill gradually the loop functions and amplitudes beyond $M_{\rm cut}$ around 1200 MeV and, with different options, we estimate uncertainties. What we observe is that, while uncertainties indeed appear in the region of $M_{\rm inv}>1200$ MeV, they are very small below that energy, rendering our predictions rather solid. The shape and strength of the mass distributions, up to a global factor (the same for all of them), are predictions of the theory which could be confronted with experiment. The ultimate aim would be to provide support to the picture in which the $f_0(500)$, $f_0(980)$ and $a_0(980)$ resonances are dynamically generated from the pseudoscalar-pseudoscalar interaction. Since neat predictions, more than reproduction of measured data, have a higher value to support one or another picture for the scalar mesons, we encourage both, calculations of the reaction in different models, as well as the performance of the reaction, which in analogy to the $\chi_{c1} \to \eta \pi^+ \pi^-$ already measured at BESIII, could be measured in this or other facilities.

\section*{Acknowledgments}

We would like to thank N. Kaiser for information concerning SU(3) invariants.
One of us, V. R. D. wishes to acknowledge the support from the Programa Santiago Grisol\'ia of Generalitat Valenciana (Exp. GRISOLIA/2015/005).
One of us, E. O. wishes to acknowledge the support from the Chinese Academy of Science in the Program of Visiting Professorship for Senior International Scientists (Grant No. 2013T2J0012).
This work is partly supported by the National Natural Science Foundation of China under Grants No. 11565007, No. 11547307 and No. 11475227. It is also supported by the Youth Innovation Promotion Association CAS (No. 2016367).
This work is also partly supported by the Spanish Ministerio de Economia y Competitividad and European FEDER funds under the contract number FIS2011-28853-C02-01, FIS2011-28853-C02-02, FIS2014-57026-REDT, FIS2014-51948-C2-1-P, and FIS2014-51948-C2-2-P, and the Generalitat Valenciana in the program Prometeo II-2014/068.

\clearpage

\bibliographystyle{plain}

\end{document}